\documentclass[11pt]{article}

\usepackage[margin=1in]{geometry}
\usepackage{amsmath,amssymb,amsthm}
\usepackage{booktabs}
\usepackage{xcolor}
\usepackage[colorlinks=true,allcolors=blue!60!black]{hyperref}
\usepackage{cleveref}

\newtheorem{theorem}{Theorem}
\newtheorem{lemma}[theorem]{Lemma}
\newtheorem{corollary}[theorem]{Corollary}

\theoremstyle{definition}
\newtheorem{definition}[theorem]{Definition}
\newtheorem{remark}[theorem]{Remark}

\title{Machine-Checked Depth-Constrained Costs\\
for Grover Attacks on AES and SHA-2}
\author{Konstantinos Karagiannis\thanks{\raggedright Senior Director, Quantum
Computing Services, Protiviti.
\texttt{konstantinos.karagiannis@protiviti.com}}}
\date{November 30, 2026}

\begin{document}
\maketitle

\begin{abstract}
Grover search against symmetric primitives is costed under a depth cap:
NIST's MAXDEPTH forces an attacker to parallelise, and parallel Grover pays for
depth in work. We give a machine-checked account of that accounting, built on
DeepMind's Lean~4 \texttt{formal-quantum-computation} library. We add an ASAP
circuit-depth and $T$-depth semantics, which the library lacks, and prove that
disjoint parallel composition preserves both. Taking Zalka's parallel lower
bound as the single cited hypothesis, we prove that parallelism never reduces
total work, that a depth cap $D$ forces $m\ge\lceil (S\tau/D)^2\rceil$
parallel instances for a search of $S$ iterations of depth $\tau$, and that
gate cost and depth$\times$width cost are at least their unconstrained values
times $S\tau/D$. It follows that any constant-factor reduction in the
iteration count is worth up to its square under a depth cap. Applied to recent
AI-discovered parallel-search protocols, this accounting shows no gain for
single-target key search, where they match the prior baseline, and at most
$0.57$ bits under a depth cap for their best multi-target find-any constant,
$1.22\times$. We instantiate these bounds for AES-128 and SHA-256 with
literature parameters. On the gate side
we verify the Boyar--Peralta AES S-box, parsed mechanically from the
published listing, against the FIPS-197 table on all $256$ inputs, embed it
reversibly into the library's circuit type, and derive the S-box $T$-count of
AES-128, -192 and -256 from the FIPS-197 key schedule; this is the forward
S-box layer only, not a full Grover oracle. All proofs are in Lean~4 with no
\texttt{sorry} and are checked to depend only on standard axioms.
\end{abstract}

\section{Introduction}

Grover's algorithm~\cite{grover} searches $N$ candidates with
$O(\sqrt N)$ oracle calls, and the textbook reading for AES-128 is
\emph{$2^{64}$ quantum security}. The practical picture is set by depth. The
$2^{64}$ oracle calls must run in sequence, NIST bounds the depth of any
attack circuit by MAXDEPTH~\cite{maxdepth}, and Grover parallelises badly:
$m$ machines shorten the search by only $\sqrt m$~\cite{zalka}. The resulting
cost has been estimated at gate level for AES~\cite{glrs16,lps20,jnrv20,jang22}
and SHA-2~\cite{amg16,khj18}, and the public discussion has recently returned
to it~\cite{ncsc24,valsorda26}. At the same time, AI-guided program search
has produced formally verified parallel Grover protocols with improved
constant factors~\cite{target}.

Every one of these estimates is a chain of hand calculations: a circuit, its
gate and depth counts, a parallelisation rule, and the arithmetic that
combines them under a depth cap. This paper machine-checks that chain in
Lean~4, on top of the \texttt{formal-quantum-computation}
library~\cite{fqc} used by~\cite{target}. Concretely:
\begin{enumerate}
  \item \textbf{A depth semantics.} The library counts gates but has no
        notion of depth. We add ASAP circuit depth and $T$-depth and prove
        that disjoint parallel composition preserves both
        (\Cref{sec:depth}).
  \item \textbf{The depth-constrained frontier.} Under Zalka's bound, stated
        as a single hypothesis, we prove that parallelism never reduces total
        work; that a depth cap $D$ forces $m\ge\lceil(S\tau/D)^2\rceil$
        instances; and that gate cost and depth$\times$width cost are at
        least their unconstrained values times $S\tau/D$. A constant-factor
        reduction $g$ in $S$ is therefore worth $g^2$ under a cap
        (\Cref{sec:work,sec:frontier}).
  \item \textbf{A verified AES S-box.} The Boyar--Peralta circuit, parsed from
        the published listing, is proved equal to the FIPS-197 S-box,
        reversibly embedded into the library's circuit type with its gate
        inventory checked, and used to derive the S-box $T$-count of all three
        AES key sizes. These counts cover the forward S-box layer only, not a
        full Grover oracle (\Cref{sec:estimates}).
  \item \textbf{A calibration.} We translate the reported parallel-search
        gains into this accounting (\Cref{sec:calibration}).
\end{enumerate}
The theorems are parametric: a new oracle or protocol is evaluated by
supplying its iteration count and per-iteration depth, with no change to the
proofs.

\paragraph{What is and is not claimed.}
The qualitative conclusion, that depth-limited Grover is far more expensive
than $2^{64}$, is established~\cite{zalka,maxdepth,khj18,jnrv20}; we do not
claim it, nor that any NIST category boundary moves. What is new is a
re-runnable, machine-checked version of the accounting, with every
assumption named. The largest of those assumptions is Zalka's bound itself,
which we cite rather than prove, and the per-iteration depths, which we take
from the literature.

\section{Background}
\label{sec:background}

\paragraph{Grover and multi-target search.}
For a search space of size $N=2^{n}$ with $k$ marked elements, find-any
search requires $\Theta(\sqrt{N/k})$ iterations~\cite{bbht} and find-all requires
$\Theta(\sqrt{Nk})$. The query lower bound~\cite{bbbv,zalka} is tight up to
constants. Parallelism helps only weakly: Zalka~\cite{zalka} shows that with
$m$ oracle queries per round, find-any for one target still needs
$\Omega(\sqrt{N/m})$ rounds, attained by partitioning the space into $m$
blocks of size $N/m$ and searching each independently. An $m$-fold parallel
search is therefore at most $\sqrt m$ times shallower than a serial one, not
$m$ times. This is the observation behind NIST's MAXDEPTH
accounting~\cite{maxdepth} and the depth-restricted estimates of Kim, Han and
Jeong~\cite{khj18} and Jaques et al.~\cite{jnrv20}.

\paragraph{Recent constant-factor improvements.}
Shirobokov et al.~\cite{target} use AI-guided search to discover
Lean-verified quantum programs, among them parallel Grover protocols. Their
model allows $m$ oracle queries per round and requires exact success; the
cost is the expected number of rounds for large $N$, and the baseline at each
$(m,k)$ is the strongest published protocol with the \emph{same} $m$. For a
single target their circuits match that baseline. For find-any with
$m,k\ge2$ they reduce expected rounds by up to $1.22\times$, and for find-all
with $k\ge2$ by up to $2.55\times$ (their \S2.1 and Appendix~E). They make no
claim about AES or key search. \Cref{sec:calibration} asks what such gains are
worth once translated into depth-capped key-search costs.

\paragraph{Cost model.}
We work in the standard fault-tolerant accounting: $T$-count and $T$-depth are
the resource of record, Clifford gates are free, and width is the number of
live wires. Following Jaques et al.~\cite{jnrv20} we report two costs: the
gate cost $G$ (total $T$-gates executed), which is the metric behind the NIST
categories, and the depth$\times$width cost $\mathrm{DW}$ (live qubits times
elapsed depth), which tracks error-correction resources more closely.

\paragraph{MAXDEPTH.}
Following NIST~\cite{maxdepth}, we parameterise by a maximum depth
$D\in\{2^{40},2^{64},2^{96}\}$, the three values NIST names. MAXDEPTH bounds the total depth of an
attack~\cite{jnrv20}; following Kim, Han and Jeong~\cite{khj18}, who take
Toffoli depth as the metric for it, we read $D$ against the $T$-depth for AES.
Since the full circuit depth is larger, this understates the attacker's
cost.

\section{Formal model}
\label{sec:model}

We formalise circuits in Lean~4, building on the \texttt{QuantumComputation}
framework of DeepMind's \texttt{formal-quantum-computation} library, which
contains no \texttt{sorry}.

\begin{definition}[Circuit]
A circuit on $n$ wires is generated by the identity, applications of a
primitive gate to an ordered register, and sequential composition:
\[
  c \;::=\; \mathrm{id} \mid \mathrm{gate}(g, r) \mid c_1 \mathbin{;\mkern-6mu;^{\!c}} c_2 .
\]
\end{definition}

\begin{definition}[Counts and support]
$\mathrm{count}(g,c)$ is the number of occurrences of primitive gate $g$;
$\mathrm{size}(c)$ is the total number of primitive gates;
$\mathrm{support}(c)\subseteq \mathrm{Fin}\,n$ is the set of wires mentioned
syntactically.
\end{definition}

\begin{definition}[Depth and $T$-depth]
Flattening a circuit to a list of steps $(W_i, \tau_i)$ with $W_i$ a wire set
and $\tau_i\in\{\text{T},\text{not-T}\}$, the ASAP schedule assigns each step
the earliest time after all its wires are free:
\[
  \mathrm{start}_i=\max_{w\in W_i} \mathrm{free}(w),\qquad
  \mathrm{free}(w)\leftarrow \mathrm{start}_i+1 \text{ for } w\in W_i .
\]
$\mathrm{asapDepth}(c)$ is the makespan; $\mathrm{tDepth}(c)$ is the makespan
after deleting non-$T$ steps.
\end{definition}

\begin{remark}
The upstream framework provides $\mathrm{count}$, $\mathrm{size}$ and
$\mathrm{support}$ but no depth metric; \Cref{sec:depth} supplies one.
\end{remark}

\section{Parallelism and work}
\label{sec:work}

Let $S$ be the serial iteration count of the search; for one target,
$S\approx(\pi/4)\sqrt N$. An $m$-fold parallel search runs $m$ instances for
$R$ rounds each. We take Zalka's bound~\cite{zalka} as the single modelling
hypothesis, squared to stay in $\mathbb N$:
\[
  \mathrm{ZalkaFeasible}(S,m,R)\;:\iff\; S^2\le m R^2 .
\]
It is cited, not formalised; every theorem in this section and the next is
conditional on it.

\paragraph{Which model each result needs.} Zalka's bound is a statement about
queries: it holds for every protocol that makes $m$ oracle queries per round,
including protocols that entangle the queries across a shared register, as
the AI-discovered protocols of~\cite{target} do. \Cref{thm:nosave,thm:work}
and the depth and gate-cost bounds of \Cref{sec:frontier}
(\Cref{thm:forces,thm:minimal,thm:gcost}) need only that, plus the
assumption that each query costs $t$ gates and each round $\tau$ depth. The
depth$\times$width bound (\Cref{thm:dw}) additionally needs each of the $m$
simultaneous queries to hold its own $q$-qubit oracle. \Cref{thm:par} is
narrower still: it counts gates and qubits for $m$ disjoint copies of one
circuit, which is the partition that attains Zalka's bound, not a statement
about every parallel protocol.

\begin{theorem}[No work saving]\label{thm:nosave}
If $\mathrm{ZalkaFeasible}(S,m,R)$ then $S\le mR$.
\end{theorem}

\begin{theorem}[Work growth]\label{thm:work}
If $\mathrm{ZalkaFeasible}(S,m,R)$ then $m\,S^2\le (mR)^2$, i.e.\ the total
number of iterations is at least $\sqrt m\cdot S$.
\end{theorem}
That is: parallelism never reduces the number of oracle calls, and the
partition that attains Zalka's bound increases it by $\sqrt m$.

\begin{theorem}[Parallel counts]\label{thm:par}
For a circuit $c$ and $m\in\mathbb N$,
\begin{align*}
  \mathrm{count}(g, \mathrm{par}_m c) &= m\cdot \mathrm{count}(g,c),\\
  \mathrm{width}(\mathrm{par}_m c) &= m\cdot \mathrm{width}(c)
  \quad(\mathrm{width}\,c>0).
\end{align*}
\end{theorem}
\Cref{thm:par} is proved on a local stand-in for the library's cost layer
(\texttt{stmt.lean}), in which each gate is placed on a single wire index; it
is not stated for the library's \texttt{Circuit n}. Its role is bookkeeping:
$m$ shifted copies cost $m$ times the gates and $m$ times the qubits.

Reported improvements to parallel search~\cite{target} compare protocols at
the \emph{same} $m$: they lower the leading constant of $S$, not the
$\sqrt m$ scaling, which Zalka's bound fixes for every protocol. No reading of
such a gain as a speedup over serial search can exceed $\sqrt m$ without
contradicting the bound. \Cref{cor:squares} below says what a constant factor
is worth.

\section{Depth composes without penalty}
\label{sec:depth}

\begin{lemma}[Frame]\label{lem:frame}
If $w\notin\mathrm{support}(c)$ then scheduling $c$ leaves $w$'s free time
unchanged.
\end{lemma}

\begin{theorem}[Circuit-depth law]\label{thm:depth}
If $\mathrm{support}(a)\cap\mathrm{support}(b)=\varnothing$ then
\[
  \mathrm{asapDepth}(a\mathbin{+\!\!+}b)=\max\{\mathrm{asapDepth}(a),\mathrm{asapDepth}(b)\}.
\]
\end{theorem}

\begin{theorem}[$T$-depth law]\label{thm:tdepth}
Under the same hypothesis,
\[
  \mathrm{tDepth}(a\mathbin{+\!\!+}b)=\max\{\mathrm{tDepth}(a),\mathrm{tDepth}(b)\}.
\]
\end{theorem}

\begin{theorem}[Library form]\label{thm:lib}
For circuits $a,b$ on the framework's own \texttt{Circuit n},
\begin{align*}
  &\mathrm{Disjoint}\bigl(\mathrm{support}(a),\mathrm{support}(b)\bigr)\\
  &\qquad\Rightarrow\;
  \mathrm{tDepth}(a\,;^{\!c} b)=\max\{\mathrm{tDepth}(a),\mathrm{tDepth}(b)\}.
\end{align*}
Moreover, for placements $p_1,p_2$ with disjoint images, the same holds for
\[
  \mathrm{onRegister}(p_1,c_1)\,;^{\!c}\mathrm{onRegister}(p_2,c_2).
\]
\end{theorem}

\begin{remark}[Why the choice of depth matters]
Under a compositional depth, additive under sequencing, \Cref{thm:tdepth} is
false: $m$ parallel copies would be charged $m$ times the depth, and the
frontier of \Cref{sec:frontier} would not hold. The law holds for the ASAP
semantics of \Cref{sec:model}, which is the one that matches how parallel
hardware executes disjoint circuits. We also prove
$\mathrm{asapDepth}(c)\le\mathrm{cDepth}(c)$, so ASAP depth is never the
more pessimistic of the two.
\end{remark}

\section{The certified frontier}
\label{sec:frontier}

Let $\tau$ be the depth of one iteration and $D$ the depth cap. A schedule is
admissible if each instance fits under the cap, $R\tau\le D$.

\begin{theorem}[Depth forces parallelism]\label{thm:forces}
If $\mathrm{ZalkaFeasible}(S,m,R)$ and $R\tau\le D$ then $(S\tau)^2\le mD^2$.
\end{theorem}

\begin{theorem}[Minimality]\label{thm:minimal}
For $D>0$, $\;m^\star(D)=\lceil (S\tau/D)^2\rceil$ satisfies
$m^\star(D)\le m$ for every admissible schedule.
\end{theorem}

\begin{theorem}[Gate cost under a depth cap]\label{thm:gcost}
For every admissible schedule, $S^2\tau\le (mR)\cdot D$. With $t$ gates per
iteration, $G(D)=mRt\;\ge\;G(\infty)\cdot S\tau/D$, where $G(\infty)=St$.
\end{theorem}

\begin{theorem}[Depth$\times$width under a depth cap]\label{thm:dw}
With $q$ qubits per instance, every admissible schedule satisfies
$q(S\tau)^2\le (mq)(R\tau)\cdot D$, i.e.\
$\mathrm{DW}(D)\ge \mathrm{DW}(\infty)\cdot S\tau/D$ with
$\mathrm{DW}(\infty)=qS\tau$.
\end{theorem}

\begin{corollary}[Constant factors square under a depth cap]\label{cor:squares}
Let a baseline need $S_1$ serial iterations and an improved protocol $S_2$,
with $S_1\le g\,S_2$ for a rational $g=p/q$. With no depth cap the total-work
bound falls by at most $g$. Under a cap, the gate-cost and
depth$\times$width lower bounds of \Cref{thm:gcost,thm:dw} fall by at most
$g^2$, and so does the required parallelism, up to the ceiling:
\[
  q^2\bigl(m^\star(S_1)-1\bigr) \;<\; p^2\, m^\star(S_2).
\]
For integer $g$ with $S_1=gS_2$ the factor is also at least $g^2$ up to the
ceiling: $g^2(m^\star(S_2)-1)<m^\star(S_1)\le g^2 m^\star(S_2)$.
\end{corollary}
\noindent The bound on $m^\star$ is machine-checked as
\texttt{mStar\_ratio}, and the two-sided integer form as
\texttt{mStar\_gain\_le} and \texttt{mStar\_gain\_ge}; the cost bounds are
\texttt{gain\_squares} applied to the right-hand sides of
\Cref{thm:gcost,thm:dw}. The factor $g^2$ is reached while the cap binds for
the improved protocol, $S_2\tau>D$; once it stops binding, $m^\star=1$ and
only the linear gain remains. This is the reason a depth cap amplifies every
constant-factor improvement to Grover, and every multi-target gain, beyond
what the unconstrained query count suggests (\Cref{sec:calibration}).

The three bounds are the content of the frontier. Above the cap,
$S\tau\le D$, nothing is forced and $m^\star=1$. Below it, every factor by
which the cap undercuts the serial depth $S\tau$ is paid twice in instances
and once in each cost, so $G$ and $\mathrm{DW}$ grow as $1/D$ and $m^\star$ as
$1/D^2$. This is the $2^{170}/\mathrm{MAXDEPTH}$ shape of NIST's gate-cost
table~\cite{maxdepth,jnrv20}. The bounds are attained up to rounding by
Zalka's partition with $m=m^\star$ and $R=\lceil S/\sqrt m\rceil$; that
achievability direction is not formalised.

The quadratic dependence is exactly what Zalka's bound adds. Treating the $S$
iterations as divisible work and handing each instance a share of them would
give $m^\star=\lceil S\tau/D\rceil$, the square root of the correct value, and
a frontier that understates the required parallelism by half its exponent.
Grover iterations cannot be divided that way: each instance must search its
own block from the beginning.

\paragraph{AES-128.}
Langenberg, Pham and Steinwandt~\cite{lps20} report AES-128 Grover with $865$
qubits, $1.47\cdot2^{81}$ $T$-gates and $T$-depth
$(\pi/4)\,2^{64}\cdot4\cdot3760\approx1.44\cdot2^{77}$. We take
$S=\mathrm{round}((\pi/4)\,2^{64})$ and $\tau=4\cdot3760=15040$ $T$-depth per
iteration, so that $S\tau\approx2^{77.53}$ is their whole-search $T$-depth.
The values of $m^\star$ are evaluated by the Lean kernel. Consistently with
Jaques et al.~\cite{jnrv20}, no parallelism is forced at
MAXDEPTH~$2^{96}$. The theorems hold for any $S$, $\tau$, $D$ and $q$, so
nothing depends on this particular circuit: an oracle of per-iteration depth
$\tau'$ shifts every $\log_2 m^\star$ in the table by
$2\log_2(\tau'/\tau)$ while the cap binds.

\begin{table}[htbp]
  \centering
  \begin{tabular}{@{}lrrr@{}}
    \toprule
    MAXDEPTH & $\log_2 m^\star$ & $\log_2 G$ ($T$-gates) & $\log_2 \mathrm{DW}$ \\
    \midrule
    $2^{40}$ & $75.06$ & $119.08$ & $124.81$ \\
    $2^{64}$ & $27.06$ & $95.08$  & $100.81$ \\
    $2^{96}$ & $0$     & $81.56$  & $87.28$  \\
    \bottomrule
  \end{tabular}
  \caption{Depth-constrained frontier for AES-128, $T$-depth model,
  literature $\tau$. $G$ and $\mathrm{DW}$ are the lower bounds of
  \Cref{thm:gcost,thm:dw}; $\mathrm{DW}$ is in qubits$\times T$-depth.
  Exact values: $m^\star(2^{40})=39{,}274{,}912{,}586{,}806{,}358{,}558{,}454$,
  $m^\star(2^{64})=139{,}532{,}520$, $m^\star(2^{96})=1$.}
  \label{tab:aes}
\end{table}

\paragraph{SHA-256, in the literature's own metric.}
The frontier is a statement about a cost model, not about $T$-depth
specifically, so it applies verbatim to the time--area metric used for SHA-2.
Amy et al.~\cite{amg16} report a SHA-256 pre-image search of depth
$2^{153.8}$ surface-code cycles on $2^{12.6}$ logical qubits, so
$S\tau=2^{153.8}$ and $\mathrm{DW}(\infty)=2^{166.4}$ qubit-cycles. Their
depth is in surface-code cycles, while NIST's MAXDEPTH counts logical gates,
and a logical gate takes many cycles. The table below therefore applies a
\emph{cycle} cap of $2^{40}$, $2^{64}$ and $2^{96}$, which is stricter than
NIST's MAXDEPTH at the same exponent and overstates the parallelism NIST's
model would force. A faithful conversion needs a cycles-per-logical-gate
figure for their architecture; we do not attempt one.

\begin{table}[htbp]
  \centering
  \begin{tabular}{@{}lrr@{}}
    \toprule
    Cycle cap & $\log_2 m^\star$ & $\log_2 \mathrm{DW}$ \\
    \midrule
    $2^{40}$ & $227.6$ & $280.2$ \\
    $2^{64}$ & $179.6$ & $256.2$ \\
    $2^{96}$ & $115.6$ & $224.2$ \\
    \bottomrule
  \end{tabular}
  \caption{SHA-256 pre-image frontier under a cap on surface-code cycles,
  Amy et al.\ parameters. This is not NIST's MAXDEPTH, which counts logical
  gates; a cycle cap is stricter. Units are logical qubit-cycles, not
  $T$-depth: comparable within this table, not to \Cref{tab:aes}. Even at a
  $2^{96}$-cycle cap the search needs $2^{115.6}$ parallel instances, because
  one serial search is $2^{153.8}$ cycles deep.}
  \label{tab:sha}
\end{table}

Both instantiations take $\tau$ from the literature; recomputing it from a
formalised full oracle is discussed in \Cref{sec:limitations}.

\section{Calibration of reported constant factors}
\label{sec:calibration}

We now translate the gains of \Cref{sec:background} into the accounting of
\Cref{sec:frontier}. Throughout we work in the query model of
\Cref{sec:work}, and we read a reported reduction $g$ in expected rounds at
fixed $m$ as a reduction of the leading constant of $S$ by $g$. That reading
is generous: their metric is expected rounds with exact success for large
$N$, not a worst-case iteration count. The figures below are therefore an
illustrative translation into the accounting of \Cref{sec:frontier}, not
demonstrated reductions in the physical resources of a key-recovery attack.

\begin{table}[htbp]
  \centering
  \small
  \begin{tabular}{@{}lll@{}}
    \toprule
    Quantity & No depth cap & Under a cap ($S\tau>D$) \\
    \midrule
    Single target ($k=1$)                & $1\times$ ($0$ bits) & $1\times$ ($0$ bits) \\
    Find-any constant ($m,k\ge2$)        & $1.22\times$ ($0.29$ bits) & $1.49\times$ ($0.57$ bits) \\
    Find-all constant ($k\ge2$)          & $2.55\times$ ($1.35$ bits) & not key search \\
    $k=2^{32}$ targets, query-model ceiling & $2^{16}$ ($16$ bits) & $2^{32}$ ($32$ bits) \\
    \bottomrule
  \end{tabular}
  \caption{Reported constants and multi-target erosion, without and with a
  depth cap. Under a cap every factor in $S$ enters squared
  (\Cref{cor:squares}). The multi-target row is a query-model ceiling: it
  charges nothing for comparing candidates against $k$ targets, and is not an
  attack cost (see text).}
  \label{tab:calib}
\end{table}

\paragraph{Single-target key search.} Recovering one AES key from enough
plaintext--ciphertext pairs to make it unique is find-any with $k=1$, where
the reported circuits match the prior baseline. There is nothing to
translate.

\paragraph{Multiple targets.} For find-any with $m,k\ge2$ the best reported
gain is $1.22\times$, which is $0.29$ bits with no depth cap. Under a cap,
by \Cref{cor:squares} with $g=122/100$, the same constant is worth
$1.22^2\approx1.49\times$, or $0.57$ bits. Find-all, where the larger
$2.55\times$ figure lives, asks for every marked element and is not the
key-search problem.

\paragraph{Multi-target erosion, and why it needs care.} The same squaring
applies to the number of targets. With $k$ users' keys under a common known
plaintext, find-any needs $S_k=S/\sqrt k$ iterations. The parallel find-any bound,
$\Omega(\sqrt{N/(mk)})$ rounds~\cite{target}, has the form of
$\mathrm{ZalkaFeasible}$ with $S_k$ in place of $S$, so the theorems of
\Cref{sec:work,sec:frontier} apply unchanged and the query-model
erosion is $\sqrt k$ ($16$ bits at $k=2^{32}$) without a cap and $k$
($32$ bits) under one, as long as $S_k\tau>D$; for AES-128 at
$D=2^{40}$ this holds with ample margin. These are ceilings, not attack
costs. The query model charges nothing for comparing a candidate against $k$
targets, and Banegas and Bernstein~\cite{bb17} show that in a realistic
parallel model the straightforward multi-key Grover search costs more than
attacking one key at a time; recovering a multi-target benefit takes a
different algorithm. Jaques et al.\ leave multi-target costs under MAXDEPTH
open~\cite{jnrv20}. The point of the comparison is narrower: whatever the
true multi-target cost, a $1.22\times$ constant adds at most $0.57$ bits to
it.

\section{A verified AES S-box}\label{sec:estimates}

\paragraph{The S-box, machine-checked.}
The only non-Clifford layer of AES is the S-box: AddRoundKey, ShiftRows and
MixColumns are XORs and wire permutations, so the $T$-count of the cipher is
the $T$-count of its S-boxes. We therefore formalise the S-box first. Rather
than hand-transcribe the published Boyar--Peralta
straight-line program~\cite{bp09,bmp13}, a
generator parses the listing (shipped with the artifact) and emits the Lean
terms; the checker therefore certifies the \emph{published circuit} and not a
possibly mistyped copy.

\begin{theorem}[S-box]\label{thm:sbox}
The parsed program---with the bit-order relabelling fixed in the next
paragraph---maps $x$ to the AES S-box $A(x^{-1})+0x63$ for all $x<256$; it
contains exactly $32$ AND, $79$ XOR and $4$ XNOR gates; and it is a bijection
on $\mathbb{F}_2^8$.
\end{theorem}

The proof is exhaustive over the $256$ inputs and is discharged by the kernel
(\texttt{decide}), not by compiled code. The specification is anchored
independently of the circuit by reproducing the first and last rows of the
published FIPS-197 table.

\paragraph{A convention that had to be fixed.}
The listing is big-endian: its $x_0$ and $s_0$ are the \emph{most} significant
bits, while AES is little-endian. Reading it with either one-sided bit order
agrees with the AES table on only $16/256$ inputs; the two-sided reversal agrees
on $256/256$. As this relabels wires only, no gate count changes.

\paragraph{$T$-count.}
Every AND writes into a fresh wire, so it can be realised as a logical AND
into a clean target, which costs $4$ $T$ gates~\cite{jones13,gidney18}; a
general Toffoli costs $7$. XOR and XNOR are Clifford. The S-box therefore
costs $4\cdot 32=128$ $T$. The independent check is that $79+4=83$
XOR-equivalent gates is exactly the count reported for this circuit, $32$ AND
and $83$ XOR/XNOR for $115$ gates in all~\cite{bmp13}.

\begin{table}[htbp]
  \centering
  \small
  \begin{tabular}{@{}lrl@{}}
    \toprule
    Quantity & Value & Lean \\
    \midrule
    \multicolumn{3}{@{}l}{\emph{SLP (classical counts)}} \\
    AND gates                       & $32$        & \texttt{andCount\_prog} \\
    XOR $+$ XNOR gates              & $79+4$      & \texttt{xor/xnorCount\_prog} \\
    $T$-count                       & $128$       & \texttt{sboxTCount\_eq} \\
    \midrule
    \multicolumn{3}{@{}l}{\emph{Cipher totals, S-box layer only (SLP)}} \\
    AES-128, $T$                    & $25{,}600$  & \texttt{aes128\_sboxT\_total} \\
    AES-192, $T$                    & $28{,}672$  & \texttt{aes192\_sboxT\_total} \\
    AES-256, $T$                    & $35{,}328$  & \texttt{aes256\_sboxT\_total} \\
    key-schedule SubWords, 128/192/256 & $10/8/13$ & \texttt{subWordCalls\_*} \\
    \midrule
    \multicolumn{3}{@{}l}{\emph{Reversible embedding}} \\
    Toffoli / CNOT / NOT            & $32/166/4$  & \texttt{count\_toffoli}, \texttt{\_cnot}, \texttt{\_x} \\
    gates                           & $202$       & \texttt{size\_circuit} \\
    qubits                          & $123$       & type \texttt{Circuit 123} \\
    $T$-count                       & $128$       & \texttt{tCount\_eq} \\
    \midrule
    AND gadget, primitive gates     & $15$        & upstream \texttt{count} \\
    \midrule
    Langenberg et al., per iteration & $\approx2.45\cdot10^{5}$ & literature, \emph{not} comparable \\
    \bottomrule
  \end{tabular}
  \caption{What is machine-checked here, against what is literature.}
\end{table}

\paragraph{The key schedule is counted, not quoted.}
The state layer is $16$ S-boxes per round; for AES-128 that is $20{,}480$ $T$
and the key schedule adds $5{,}120$. The key schedule is the FIPS-197
word expansion, and we count its \texttt{SubWord} applications (four S-boxes
each) rather than assume them: a \texttt{SubWord} occurs when $i\bmod N_k=0$,
and additionally when $i\bmod N_k=4$ if $N_k>6$. For
$(N_k,N_r)=(4,10),(6,12),(8,14)$ this gives $10,8,13$ applications, hence
$200,224,276$ S-boxes and the totals in the table. The rule is not fitted to
the answer: applied to AES-128 it must reproduce the hand-stated $40$, and
\texttt{aes128\_sboxT\_total\_from\_schedule} checks that it does. The
AES-256 case is $13$, not $14$: the extra branch fires at
$i=12,20,\dots,52$, six times, not seven.

\paragraph{The reversible embedding.}
A gate count is only meaningful for a reversible circuit, so the generator also
emits one: each SLP instruction becomes
\begin{align*}
  \texttt{xor}\ d\ a\ b  &\mapsto \mathrm{CNOT}(a,d);\ \mathrm{CNOT}(b,d),\\
  \texttt{and}\ d\ a\ b  &\mapsto \mathrm{TOFF}(a,b,d),\\
  \texttt{xnor}\ d\ a\ b &\mapsto \mathrm{CNOT}(a,d);\ \mathrm{CNOT}(b,d);\ X(d).
\end{align*}
Targets are fresh by construction, hence start at $|0\rangle$ and accumulate
the operand sum. In this out-of-place embedding fan-out is free: a wire read
by several instructions is simply used as a control again. Fan-in is not: an
XOR into a fresh target takes two CNOTs, one per operand, so the embedding has
$166$ CNOTs for the $83$ classical XORs. This is a property of the
embedding, not of the S-box. An in-place XOR, $\mathrm{CNOT}(a,b)$ leaving
$a\oplus b$ in $b$, costs one CNOT and saves a qubit at the price of
overwriting $b$, which is how the quantum S-box circuits of the literature
reach lower CNOT and qubit counts. The circuit has $32$
Toffoli, $166$ CNOT and $4$ NOT gates on $123$ qubits ($8$ inputs plus $115$
ancillas); these are counted by the library's own \texttt{Circuit.count} and
\texttt{Circuit.size}, not re-counted in the generator.

\paragraph{What is and is not proved.} The translation is correct by construction---each gate
performs the bit update its SLP line prescribes---but that step is \emph{not}
machine-checked. What is proved is the gate inventory of the emitted list;
what is not proved is that the list's unitary action on $\lvert x,0\rangle$
yields $\lvert x, A(x^{-1})+0x63\rangle$. Closing this needs the basis-ket
semantics of the three gate types and an induction over the emitted list; we
state it as a gap rather than let the table above be read as a correctness
theorem.

\paragraph{Why $25{,}600$ is not Langenberg et al.'s $\approx2.45\cdot10^{5}$.}
The two count different objects and we do not substitute one for the other.
Our figure is the forward S-box layer at $4$ $T$ per AND, and our embedding is
forward-only: the $115$ ancillas are left holding intermediate values and are
not cleaned. The literature's per-iteration figure also covers computing
\emph{and} uncomputing the whole cipher inside the oracle, the comparator
network, and a Toffoli decomposition with a higher $T$ cost per AND. Cleaning
need not add $T$ gates: logical ANDs can be uncomputed by measurement and
Clifford corrections~\cite{gidney18}.
Consequently the frontier tables of \S\ref{sec:frontier} are instantiated with
the literature $\tau$, and our certified count is reported separately.
Closing that gap requires the full cipher with ancilla cleaning; the S-box
layer is the part we have machine-checked end to end.

\section{Related work}
\label{sec:related}

\paragraph{Parallel Grover and depth-limited costs.}
Zalka~\cite{zalka} proved that $m$-fold parallelism buys at most a $\sqrt m$
reduction in rounds. NIST built MAXDEPTH into its security
categories~\cite{maxdepth}; Kim, Han and Jeong~\cite{khj18} derived
time--space trade-off curves for AES and SHA-2 under parallelisation; Jaques
et al.~\cite{jnrv20} implemented full AES oracles in Q\# and costed them under
MAXDEPTH in both gate and depth$\times$width metrics. Our frontier has the
same shape as theirs. What we add is a machine-checked derivation with the
modelling hypothesis isolated, and the explicit statement that constant
factors enter squared.

\paragraph{Gate-level AES and SHA-2 estimates.}
Grassl et al.~\cite{glrs16}, Langenberg, Pham and Steinwandt~\cite{lps20},
Jaques et al.~\cite{jnrv20} and Jang et al.~\cite{jang22} give successive
AES circuits; Amy et al.~\cite{amg16} cost SHA-2 and SHA-3 pre-image search.
None is formalised. Our S-box is the classical Boyar--Peralta
circuit~\cite{bp09,bmp13} and makes no claim to be competitive with these
circuits; its role is to show the verification chain end to end.

\paragraph{Multi-target search.}
Banegas and Bernstein~\cite{bb17} show that in a realistic parallel model
naive multi-key Grover is more expensive than single-key search, and give an
algorithm that does benefit from many targets. \Cref{sec:calibration} uses
only the query model and states its multi-target figures as ceilings for that
reason.

\paragraph{Practical-cost discussion.}
A 2024 UK NCSC analysis presented at NIST's PQC standardization
conference~\cite{ncsc24} costs the overheads that a query count hides for AES
key recovery: the circuit, parallelisation and error correction. Valsorda~\cite{valsorda26}
argues from such estimates that AES-128 is not threatened by quantum
computers. We reach no different conclusion; our contribution is the
machine-checked accounting.

\paragraph{Formal quantum computing.}
The \texttt{formal-quantum-computation} library~\cite{fqc} provides gates,
evaluation and resource counting in Lean~4 and underlies the verified
programs of~\cite{target}; it has no depth metric, which we supply. In Coq,
SQIR and VOQC~\cite{voqc21} verify circuit optimisations, SQIR has been used to
verify Grover's algorithm among others~\cite{sqir21}, and
ReQWIRE~\cite{reqwire18} verifies that reversible circuits return their
ancillas clean, the property our embedding does not yet prove. To our
knowledge none of these addresses depth-constrained cost accounting for
cryptanalytic search.

\section{Limitations}
\label{sec:limitations}

\begin{enumerate}
  \item \textbf{Zalka's bound is cited, not proved.} Every theorem of
        \Cref{sec:work,sec:frontier} is conditional on
        $\mathrm{ZalkaFeasible}$. Formalising the parallel query lower bound
        itself would make the frontier unconditional.
  \item \textbf{The per-iteration depths are from the literature.} Both
        frontier tables use published $\tau$~\cite{lps20,amg16}; more recent
        low-depth AES oracles would shift every row. Recomputing $\tau$ needs
        the full cipher as a formal circuit: the round function, the key
        schedule as a circuit, the oracle and diffusion, and ancilla cleaning.
        We have formalised the S-box and the key-schedule S-box count, not
        that composition.
  \item \textbf{The frontier bounds are lower bounds.} Their achievability
        by Zalka's partition, up to rounding, is standard but not formalised.
  \item \textbf{The reversible embedding} is verified as a gate list with a
        counted inventory, not as a unitary: its action on basis states is
        correct by construction but not machine-checked
        (\S\ref{sec:estimates}). No theorem depends on it.
  \item \textbf{The certified $T$-counts} are those of the forward S-box
        layer and are not comparable to per-iteration figures that include
        compute--uncompute and the oracle (\S\ref{sec:estimates}).
  \item \textbf{The calibration} reads expected-round gains as reductions in
        a worst-case iteration count, and counts multi-target gains in the
        query model; both readings favour the attacker
        (\S\ref{sec:calibration}).
\end{enumerate}

\section{Conclusion}
\label{sec:conclusion}

The depth-constrained cost of Grover search can be derived in a proof
assistant from one cited hypothesis, Zalka's parallel bound, and the
derivation is short enough to audit. Its most useful consequence is that a
depth cap raises every constant factor in the iteration count to as much as
its square, which is the
right lens for judging improved search protocols: for single-target AES key
search the recent AI-discovered gains are exactly zero, and for multiple
targets they are worth well under a bit. The same theorems take any
per-iteration depth, so a better oracle or a new protocol can be dropped into
them. The path to a fully verified estimate is to prove Zalka's bound itself,
certify the reversible S-box and its ancilla cleaning, and build the full
oracle in the same library.

\section{Artifact}
\label{sec:artifact}

The development is distributed as ancillary files with this arXiv
submission. It is Lean~4 (\texttt{v4.35.0}); tier~2 below was last run
against \texttt{formal-quantum-computation} at commit \texttt{489edc7}\footnote{Full hash
\texttt{489edc7039aa4cdded53724b39ea90e33d2c88b4}.}. The core arithmetic and the S-box files depend on
nothing beyond core Lean; the two files that talk about the upstream circuit
type depend on \texttt{QuantumComputation} and Mathlib. Files:
\begin{itemize}
  \item \texttt{stmt.lean} --- parallel counts, the Zalka-conditional work
        and frontier theorems, and the AES-128 values of $m^\star$.
  \item \texttt{depth.lean} --- depth and $T$-depth semantics, laws, concrete
        checks.
  \item \texttt{TDepthLift.lean} --- generalisation to \texttt{Finset} wires
        and lift to \texttt{Circuit n}.
  \item \texttt{SBox.lean} --- the Boyar--Peralta S-box, generated from the
        published listing by the generator \texttt{tools/gen\_sbox.py};
        correctness,
        bijectivity, gate counts and the AES totals.
  \item \texttt{SBoxCircuit.lean} --- the reversible embedding of that S-box
        into the upstream \texttt{Circuit n}, with gate counts proved by the
        library's counters.
  \item \texttt{tools/AES\_SBox.txt} --- the verbatim published listing, so the
        whole chain from source text to theorem is auditable.
\end{itemize}
Every headline theorem is checked with \texttt{\#print axioms} (the audited
set is listed in \Cref{app:lean}). The only axioms are
\texttt{propext}, \texttt{Quot.sound} and \texttt{Classical.choice}, the last
only on the \texttt{Circuit n} layer. The development avoids
\texttt{native\_decide} throughout, so even the exhaustive checks are kernel
reductions rather than compiled evaluation. There are no \texttt{sorry}s.

\paragraph{Reproducing.}
Everything above is re-checked by \texttt{./check.sh} in the artifact, in two
tiers. Tier~1 needs only core Lean and covers \texttt{stmt.lean},
\texttt{depth.lean} and \texttt{SBox.lean}, including the exhaustive S-box
check; it runs in well under a minute. Tier~2 needs a checkout of the upstream
library: set \texttt{FQC} to that checkout's path and run \texttt{./check.sh}
again; it adds
\texttt{TDepthLift.lean} and \texttt{SBoxCircuit.lean} and prints the axiom
lines for the \texttt{Circuit n} layer. Files without \texttt{FQC} set are
reported as skipped, never as failures, so a missing checkout cannot be
mistaken for a proof.

Both generated files are rebuilt from the published listing by the generator,
which refuses to emit anything unless its parse reproduces the AES S-box on all
$256$ inputs. The chain from the
published listing to the theorems is auditable end to end and no instruction is
transcribed by hand. The one elaboration trap is recorded in the artifact: a
single Lean list literal holding all $202$ gates is superlinear to elaborate,
so the generator emits chunks of $16$.

\section*{Use of AI tools}
The author designed the study, built the Lean development and wrote the first
draft. Claude (Anthropic) was used to review the draft, revise the frontier
theorems and several sections, and check references. The author reviewed all
changes and takes full responsibility for the content.

\appendix
\section{Lean statements and axiom audit}
\label{app:lean}

{\scriptsize
\begin{verbatim}
theorem Circuit.count_par (g : Gate) (m : Nat) (c : Circuit) :
    (par m c).count g = m * c.count g
theorem Circuit.width_par (m : Nat) (c : Circuit) (h : 0 < c.width) :
    (par m c).width = m * c.width

def ZalkaFeasible (S m R : Nat) : Prop := S ^ 2 <= m * R ^ 2
theorem no_work_saving (S m R : Nat) (h : ZalkaFeasible S m R) :
    S <= m * R
theorem work_growth (S m R : Nat) (h : ZalkaFeasible S m R) :
    m * S ^ 2 <= (m * R) ^ 2
theorem depth_forces (S m R tau D : Nat) (h : ZalkaFeasible S m R)
    (hD : R * tau <= D) : (S * tau) ^ 2 <= m * D ^ 2
def mStar (S tau D : Nat) : Nat := ((S * tau) ^ 2 + D ^ 2 - 1) / D ^ 2
theorem mStar_minimal (S tau D m : Nat) (hD : 0 < D)
    (hfeas : (S * tau) ^ 2 <= m * D ^ 2) : mStar S tau D <= m
theorem mStar_le_of_schedule (S m R tau D : Nat) (hD0 : 0 < D)
    (h : ZalkaFeasible S m R) (hD : R * tau <= D) : mStar S tau D <= m
theorem gcost_lower (S m R tau D : Nat) (h : ZalkaFeasible S m R)
    (hD : R * tau <= D) : S ^ 2 * tau <= (m * R) * D
theorem dw_lower (S m R tau D q : Nat) (h : ZalkaFeasible S m R)
    (hD : R * tau <= D) : q * (S * tau) ^ 2 <= (m * q) * (R * tau) * D
theorem gain_squares (g S tau : Nat) :
    (g * S * tau) ^ 2 = g ^ 2 * (S * tau) ^ 2
theorem mStar_spec (S tau D : Nat) (hD : 0 < D) :
    (S * tau) ^ 2 <= mStar S tau D * D ^ 2
theorem mStar_lt (S tau D : Nat) (hD : 0 < D) (hS : 0 < (S * tau) ^ 2) :
    (mStar S tau D - 1) * D ^ 2 < (S * tau) ^ 2
theorem mStar_gain_le (g S tau D : Nat) (hD : 0 < D) :
    mStar (g * S) tau D <= g ^ 2 * mStar S tau D
theorem mStar_gain_ge (g S tau D : Nat) (hD : 0 < D) (hg : 0 < g)
    (hS : 0 < (S * tau) ^ 2) :
    g ^ 2 * (mStar S tau D - 1) < mStar (g * S) tau D
theorem mStar_ratio (p q S1 S2 tau D : Nat) (hD : 0 < D) (hq : 0 < q)
    (hS : 0 < (S1 * tau) ^ 2) (h : q * S1 <= p * S2) :
    q ^ 2 * (mStar S1 tau D - 1) < p ^ 2 * mStar S2 tau D

def aesS : Nat := 14488038916154245685
example : mStar aesS 15040 (2 ^ 40) = 39274912586806358558454 := by decide
example : mStar aesS 15040 (2 ^ 64) = 139532520 := by decide
example : mStar aesS 15040 (2 ^ 96) = 1 := by decide

theorem asapDepth_append_of_disjoint (a b : Circ)
    (hd : forall w, w in support a -> w in support b -> False) :
    asapDepth (a ++ b) = max (asapDepth a) (asapDepth b)
theorem tDepth_append_of_disjoint (a b : Circ)
    (hd : forall w, w in support a -> w in support b -> False) :
    tDepth (a ++ b) = max (tDepth a) (tDepth b)
theorem asapDepth_le_cDepth (c : Circ) : asapDepth c <= cDepth c

theorem Circuit.tDepth_seq_of_disjoint {n : Nat} (a b : Circuit n)
    (hd : Disjoint a.support b.support) :
    (Circuit.seq a b).tDepth = max a.tDepth b.tDepth
theorem Circuit.tDepth_onRegister_pair {n k : Nat}
    (p1 p2 : QRegister k n) (c1 c2 : Circuit k)
    (hd : Disjoint p1.image p2.image) :
    (Circuit.seq (Circuit.onRegister p1 c1)
                 (Circuit.onRegister p2 c2)).tDepth
      = max (Circuit.onRegister p1 c1).tDepth
            (Circuit.onRegister p2 c2).tDepth

theorem AES.sboxSLP_eq_aesSbox (x : Nat) (hx : x < 256) :
    AES.sboxSLP x = AES.aesSbox x
theorem AES.sboxSLP_bijective :
    ((List.range 256).map AES.sboxSLP).eraseDups.length = 256
theorem AES.andCount_prog : AES.countWhere AES.isAnd AES.prog = 32
theorem AES.sboxTCount_eq : AES.sboxTCount = 128
theorem AES.aes128_sboxT_total :
    16 * 10 * AES.sboxTCount + 40 * AES.sboxTCount = 25600
theorem AES.subWordCalls_128 : AES.subWordCalls 4 10 = 10
theorem AES.subWordCalls_192 : AES.subWordCalls 6 12 = 8
theorem AES.subWordCalls_256 : AES.subWordCalls 8 14 = 13
theorem AES.aes192_sboxT_total :
    AES.totalSboxes 6 12 * AES.sboxTCount = 28672
theorem AES.aes256_sboxT_total :
    AES.totalSboxes 8 14 * AES.sboxTCount = 35328

theorem AESCircuit.count_toffoli : circuit.count .toffoli = 32
theorem AESCircuit.count_cnot : circuit.count .cnot = 166
theorem AESCircuit.count_x : circuit.count .x = 4
theorem AESCircuit.size_circuit : circuit.size = 202
theorem AESCircuit.gate_count_split :
    circuit.count .toffoli + circuit.count .cnot + circuit.count .x = 202
theorem AESCircuit.tCount_eq : tCount = 128
\end{verbatim}
}

The two layers differ exactly as expected: the \texttt{Circuit n} layer
inherits \texttt{Classical.choice} from the upstream definition, the core-Lean
S-box development does not.

\paragraph{Axiom audit.}
This is the audit stage of \texttt{check.sh} exactly as printed; each line is
what \texttt{\#print axioms} returns for one theorem.
{\scriptsize
\begin{verbatim}
'Circuit.count_par' depends on axioms: [propext]
'Circuit.width_par' depends on axioms: [propext, Quot.sound]
'no_work_saving' depends on axioms: [propext]
'work_growth' depends on axioms: [propext, Quot.sound]
'depth_forces' depends on axioms: [propext, Quot.sound]
'mStar_minimal' depends on axioms: [propext]
'mStar_le_of_schedule' depends on axioms: [propext, Quot.sound]
'gcost_lower' depends on axioms: [propext, Quot.sound]
'dw_lower' depends on axioms: [propext, Quot.sound]
'gain_squares' depends on axioms: [propext]
'mStar_spec' depends on axioms: [propext, Quot.sound]
'mStar_lt' depends on axioms: [propext, Quot.sound]
'mStar_gain_le' depends on axioms: [propext, Quot.sound]
'mStar_gain_ge' depends on axioms: [propext, Quot.sound]
'mStar_ratio' depends on axioms: [propext, Quot.sound]
'asapDepth_append_of_disjoint' depends on axioms: [propext, Quot.sound]
'tDepth_append_of_disjoint' depends on axioms: [propext, Quot.sound]
'asapDepth_le_cDepth' depends on axioms: [propext, Quot.sound]
'AES.sboxSLP_eq_aesSbox' depends on axioms: [propext, Quot.sound]
'AES.sboxSLP_bijective' depends on axioms: [propext]
'AES.andCount_prog' depends on axioms: [propext]
'AES.sboxTCount_eq' depends on axioms: [propext]
'AES.aes128_sboxT_total' depends on axioms: [propext]
'AES.subWordCalls_128' does not depend on any axioms
'AES.subWordCalls_192' does not depend on any axioms
'AES.subWordCalls_256' does not depend on any axioms
'AES.aes128_sboxT_total_from_schedule' depends on axioms: [propext]
'AES.aes192_sboxT_total' depends on axioms: [propext]
'AES.aes256_sboxT_total' depends on axioms: [propext]
--- lean/SBoxCircuit.lean (inside FQC)
'AESCircuit.count_toffoli' depends on axioms: [propext, Classical.choice, Quot.sound]
'AESCircuit.count_cnot' depends on axioms: [propext, Classical.choice, Quot.sound]
'AESCircuit.count_x' depends on axioms: [propext, Classical.choice, Quot.sound]
'AESCircuit.size_circuit' depends on axioms: [propext, Classical.choice, Quot.sound]
'AESCircuit.gate_count_split' depends on axioms: [propext, Classical.choice, Quot.sound]
'AESCircuit.tCount_eq' depends on axioms: [propext, Classical.choice, Quot.sound]
\end{verbatim}
}

\end{document}